\documentclass[11pt]{article}
\title{Mechanism for Resolving Gauge Hierarchy\\and Large Vacuum Energy}
\author{Robert D. Klauber\cite{RDKlauber}}
\date{Original: Febrary 6, 2018\\  Minor revision, including title change: March 4, 2019}

\usepackage {graphicx}
\usepackage {longtable}
\usepackage{changepage}
\usepackage{hyperref} 
\begin{document}
\maketitle

\begin{abstract}
Alternative forms of the solutions to the quantum field equations and their 
implications for physical theory are considered. Incorporation of these 
alternative solution forms, herein deemed ``supplemental solutions'', into the 
development of quantum field theory leads to a unique class of 
particle states, which  
may provide simple resolutions of more than 
one extant problem in high energy physics. The symmetry between the 
traditional and supplemental solutions results in a direct and natural 
zero-point energy value of zero, and, as well, a possible mechanism for 
cancelling the Higgs condensate energy, thereby providing a potential 
resolution of the large cosmological constant problem. Further, this 
symmetry may also resolve the Higgs gauge hierarchy problem. Resolutions of 
seeming theoretic impediments to supplemental fields, in particular, 
non-positive definite Fock space metric and vacuum decay, are presented, and concomitant implications for 
unitarity are considered. As supplemental solutions are 
already inherent in quantum field theory, little change is required to the 
fundamental mathematics of the theory.

\end{abstract}

\section{Introduction}
\label{intro}
\begin{adjustwidth}{.4in}{.4in}
\textit{``\textellipsis given that everything turned out to be very simple }[\textit{experimentally}]\textit{, yet extremely puzzling -- puzzling in its simplicity ... We have to \textellipsis find the new principles that will explain the simplicity.''}
\end{adjustwidth}
\hspace{200pt}
Neil Turok

\subsection{Unresolved Issues in Physics}
\label{subsec:Unresolved}
As summarized by Peebles and Ratra\cite{Peebles:2003}, 
Padmanabhan\cite{Padmanabhan:2003}, and others, there are presently three 
overriding cosmologic issues involving observed phenomena for which no 
generally accepted theoretical solutions exist: 1) dark matter 
(non-baryonic, unseen ``normal'' matter), 2) dark energy (with effective 
negative pressure and positive mass-energy), and 3) a vanishing sum of 
zero-point energies (ZPE). These may be related, or unrelated. 

Additionally, with the increasing lack of experimental support for 
supersymmetry (SUSY), the Higgs gauge hierarchy problem has grown more 
intractable.

The ZPE could give rise, in a manner suggested by Zeldovich\cite{Zel:1968}, 
and further refined by Martin\cite{Martin:2012}, to a cosmological 
constant, and thus, there are two aspects of what is known as the 
cosmological constant (c.c.) problem. First, why is vacuum energy not 
enormous (``the large c.c. problem''), and second, why is it almost null, 
but not completely null (``the small c.c. problem''). 

This article focuses on a possible resolution of the large c.c. and gauge 
hierarchy problems via incorporation into quantum field theory of 
inherent, but heretofore apparently unappreciated, alternative forms for the solutions
to the field equations. Though possibly 
seeming, at first, to be similar to other approaches to one or both of these 
issues presented, or summarized, in Nobbenhuis\cite{Nobbenhuis:2006}, 
Sakharov\cite{Sakharov:2000}, 
Linde\cite{Linde:1984}\cite{Linde:1988}, 
Chen\cite{Chen:1}\cite{Chen:2}, 
Henry-Couannier\cite{Henry:1}\cite{Henry:2005}, 
Faraoni\cite{Faraoni:2004}, Quiros\cite{Quiros:2004}, D'Agostini et 
al\cite{Agostini:2005}, Habara et al\cite{Habara:2007}, Kaplan and 
Sundrum\cite{Kaplan:2006}, 
Moffat\cite{Moffat:2005}\cite{Moffat:2006}, Elze\cite{Elze:2007}, 
't Hooft and Nobbenhuis\cite{Hooft:2006}, Andrianov et 
al\cite{Andrianov:2007}, and Kawamura\cite{Kawamura:2015}, the approach shown herein is, at its foundation, 
fundamentally different.

\subsection{Negative Frequencies and Supplemental Solutions}
\label{subsec:negative}
The issue of negative-frequency solutions to the relativistic counterparts 
of the Schr\"{o}dinger equation has a long and variegated history. Such 
solutions constitute a second way to solve the quantum field equations 
beyond those of positive frequency. The question of interpretation of 
negative-frequency solutions, one of the most famous in the history of 
science, was answered via field (second) quantization. The solutions to the 
field equations could then be shown to be operators that create and destroy 
(positive energy) states, rather than being states themselves.

By including the negative-frequency solutions, the set of solutions to the 
quantum field equations was doubled in size, and thus so were the number of 
particles. In this article, it is noted that alternative solution forms to 
the field equations, interpreted within a framework of physical law, double 
the number of particle states once again; i.e. they introduce new particles. 
The typically unused set of solution forms is designated herein as the set 
of ``\textit{supplemental }solutions'', and the new particles as ``supplemental particles''.

\section{Alternative Solutions to the Field Equations}
\label{alternative}
For simplicity, we focus on the scalar field equation (in natural units)
\begin{equation}
\label{eq1}
\left( {\partial_{\mu } \partial^{\mu }+\mu^{2}} \right)\phi =0
\end{equation}
with traditional eigensolutions
\begin{equation}
\label{eq2}
e^{-ikx}=e^{-i{\omega t+i{\rm {\bf k}}\cdot {\rm {\bf x}}} },
\end{equation}
\begin{equation}
\label{eq3}
e^{ikx}=e^{i{\omega t-i{\rm {\bf k}}\cdot {\rm {\bf x}}} },
\end{equation}
Note, however, that by taking $\omega \to -\omega $ (consider the 
symbol $\omega $ as always a positive number) in (\ref{eq2}) and (\ref{eq3}), one obtains 
alternative solution forms for (\ref{eq1})\cite{Klauber:2007} of
\begin{equation}
\label{eq4}
e^{-i\underline{k}x}=e^{i{\omega t+i{\rm {\bf k}}\cdot {\rm {\bf x}}} },
\end{equation}
\begin{equation}
\label{eq5}
e^{i\underline{k}x}=e^{-i{\omega t-i{\rm {\bf k}}\cdot {\rm {\bf x}}} },
\end{equation}
New notation (see underscoring above) is herein introduced, wherein (\ref{eq4}) and 
(\ref{eq5}) are the aforementioned supplemental solutions to (\ref{eq1}). Supplemental
solutions are not mathematically independent from the traditional solutions.
(See the appendix for details.) Nevertheless, when used to develop quantum field
theory (QFT) in the conventional manner, they give rise to operators that
behave differently. In fact, 
they give rise to distinctly different physical properties for particles. This is 
shown explicitly in the appendix for relativistic quantum mechanics (RQM) 
and derived later in this article for QFT.

That is, if one simply 
starts with (\ref{eq4}) and (\ref{eq5}), and proceeds (as we do in this article) through the 
same steps used to develop traditional QFT, one finds 
states created and destroyed by supplemental field operators that have 
distinctly different eigenvalues (quantum numbers) from traditional particle 
states, and thus comprise an independent species of particle.

Note that similar results could be attained by using the traditional solution forms 
(\ref{eq2}) and (\ref{eq3}) with different assumptions for commutators and certain operators. 
But that approach seems decidedly more \textit{ad hoc}, is less tidy and natural, and obscures 
the parallels between states in RQM and QFT. 

Note also that ${\pm ({\omega t+{\rm {\bf k}}\cdot {\rm {\bf x}}}) }$ is Lorentz invariant, as it is simply an expression of the world scalar $\mp k_{\mu } x^{\mu }$ with a different value for $k_0$.

The Dirac and Proca equations are also solved by supplemental solutions forms. As shown in Klauber\cite{Klauber:1}, the Dirac equation actually 
has eight eigenspinor solutions, a set of four for $e^{+(iEt-i{\rm {\bf 
p}}\,\cdot {\rm {\bf x}})}$ and another set of four for $e^{-(iEt-i{\rm {\bf 
p}}\,\cdot {\rm {\bf x}})}$, as one would expect in solving a 4X4 matrix 
eigenvalue problem. Four (two from the first set and two from the second 
set) have spacetime dependence like that of (\ref{eq2}) and (\ref{eq3}) and are employed in 
traditional QFT. Four (the other two from the first set and the other two 
from the second set) have spacetime dependence like that of (\ref{eq4}) and (\ref{eq5}) and 
are not employed in traditional QFT.

In RQM, the precursor to 
QFT\cite{Klauber:2013}, (\ref{eq3}) presented a problem as it represented a 
negative-energy state. As noted, QFT resolved this, but examination of (\ref{eq4}) 
and (\ref{eq5}) in the context of RQM leads to a similar issue of negative energy, 
as well as an additional one. Momentum direction in (\ref{eq4}) and (\ref{eq5}), if they 
represent physical states, is in the opposite direction of wave velocity 
[see the appendix], and hence (\ref{eq4}) and (\ref{eq5}) are unlikely candidates for 
physical particle states. Possibly for this reason, solutions of the form of 
(\ref{eq4}) and (\ref{eq5}) were not used in the development of RQM.

In parallel with the historical development of (\ref{eq2}) and (\ref{eq3}) in QFT, however, 
we can apply second quantization to (\ref{eq4}) and (\ref{eq5}), determine the resultant 
field operator solutions, and see if those solutions might provide anything 
of value in helping to match experiment with theory. 

\section{Supplemental Solutions and QFT}
\label{sec:supplemental}
\subsection{Symmetry of Solution Forms}
If (\ref{eq4}) and (\ref{eq5}) solve the same free field equations as (\ref{eq2}) and (\ref{eq3}), then it 
follows that the free scalar Lagrangian is symmetric under the 
transformation $\omega \to -\omega $. However, the full traditional Lagrangian does 
not have such symmetry under change of sign of energy for all fields.

Symmetry, as used herein, refers to the supplemental solutions being a “mirror” image 
of the traditional free field solutions, and implies that changing the sign of energy in those traditional solutions results in alternative (supplemental) solutions.

\subsection{Klein-Gordon Supplemental Solutions}
\label{subsec:Klein-Gordon}
Quantum field theory formalism for the supplemental solutions can be 
developed in parallel with the standard approach, using, for scalar fields, 
the following definitions.
\begin{equation}
\label{eq6}
\underline{\phi }\underline{}=\underline{\phi 
}^{+}\underline{}+\underline{\phi }^{-}=\sum\limits_{\mbox{k}} {\left( 
{\frac{1}{2V\omega_{k} }} \right)^{1/2}\left\{ {\underline{a}({\rm {\bf 
k}})e^{-i\underline{k}x}+\underline{b}^{\dag }({\rm {\bf 
k}})e^{\underline{k}x}} \right\}} 
\end{equation}
\begin{equation}
\label{eq7}
\underline{}\underline{\phi }^{\dag }=\underline{\phi }^{\dag 
+}+\underline{\phi }^{\dag -}=\sum\limits_{\mbox{k}} {\left( 
{\frac{1}{2V\omega_{k} }} \right)^{1/2}\left\{ {\underline{b}({\rm {\bf 
k}})e^{-i\underline{k}x}+\underline{a}^{\dag }({\rm {\bf 
k}})e^{i\underline{k}x}} \right\}} 
\end{equation}
\begin{equation}
\label{eq8}
\underline{\cal L}_{0}^{0} =\partial_{\mu } \underline{\phi }^{\dag }\partial 
^{\mu }\underline{\phi }\underline{}-\mu^{2}\underline{\phi }^{\dag 
}\underline{\phi }\underline{},
\end{equation}
where (\ref{eq8}) is an extra component added to the Lagrangian density representing 
the scalar supplemental solutions. The superscript ``0'' refers to ``free'' 
Lagrangian; the subscript ``0'', to scalar fields.

\subsection{Quantizing Supplemental Solutions}
\label{subsec:quantizing}
Applying second quantization to the supplemental solutions whereby we take classical Poisson brackets over to quantum commutators (see ref. \cite{Klauber:2013}. pp. 52-53),
\begin{equation}
\label{eq9}
\left[ {\underline{\phi_{r} }({\rm {\bf x}},t),\underline{\pi_{s} }({\rm 
{\bf y}},t)} \right]=\left[ {\underline{\phi_{r} }({\rm {\bf 
x}},t),\underline{\dot{{\phi }}_{s} }^{\dag }({\rm {\bf y}},t)} 
\right]=i\delta_{rs} \delta ({\rm {\bf x}}-{\rm {\bf 
y}})\,\,\,\,\,\,
\end{equation}
and with (\ref{eq6}) and (\ref{eq7}), this yields the coefficient commutation 
relations\cite{Klauber:1} 
\begin{equation}
\label{eq10}
\left[ {\underline{a}\left( {{\rm {\bf k}}} 
\right),\underline{a}^{\mbox{\dag }}\left( {{\rm {\bf {k}'}}} \right)} 
\right]=\left[ {\underline{b}_{r} \left( {{\rm {\bf k}}} 
\right),\underline{b}\left( {{\rm {\bf {k}'}}} \right)} \right]=-\delta 
_{{\rm {\bf k{k}'}}} 
\end{equation}
Note the above relations differ from their traditional solution counterparts 
by the minus sign on the RHS. This resulted from the time derivative in (\ref{eq9}), 
since the supplemental solutions have opposite signs from the traditional 
solutions for the time (energy) term in the exponent.

\subsection{The Scalar Supplemental Hamiltonian}
\label{subsec:mylabel1}
Using (\ref{eq6}) and (\ref{eq7}) in the relevant term in the Hamiltonian density
\begin{equation}
\label{eq11}
\underline{{\cal H}}_{0}^{0} =\sum\limits_r {\underline{\pi_{r} 
}\underline{\dot{{\phi }}_{r} }^{\mbox{\dag }}-\underline{{\cal 
L}}_{0}^{\,0} } =\underline{\dot{{\phi }}}\underline{\dot{{\phi 
}}}^{\mbox{\dag }}+\underline{\dot{{\phi }}}^{\mbox{\dag 
}}\underline{\dot{{\phi }}}-\underline{{\cal L}}_{0}^{\,0} 
=\underline{\dot{{\phi }}}\underline{\dot{{\phi }}}^{\mbox{\dag }}+\nabla 
\underline{\phi }^{\dag }\nabla \underline{\phi }+m^{2}\underline{\phi 
}^{\dag }\underline{\phi }\underline{}\underline{}
\end{equation}

\noindent
and integrating over all space, in parallel with that of ref. 
\cite{Klauber:2013}, pp. 53-54, for the traditional fields, yields 
\begin{equation}
\label{eq12}
\underline{H}=\sum\limits_{{\rm {\bf k}}} {\omega_{{\rm {\bf k}}} } \left\{ 
{\underline{a}^{\dag }(\mathbf{k})\underline{a}(\mathbf{k})-\textstyle{1 \over 
2}+\underline{b}^{\dag }(\mathbf{k})\underline{b}(\mathbf{k})-\textstyle{1 
\over 2}} \right\}
\end{equation}
where for notational streamlining we here and from henceforth drop the sub 
and superscript notation. Note that, due to (\ref{eq10}), the 1/2 terms in (\ref{eq12}) have 
the opposite sign from similar terms in the traditional Hamiltonian (\ref{eq13}),
\begin{equation}
\label{eq13}
H=\sum\limits_{{\rm {\bf k}}} {\omega_{{\rm {\bf k}}} } \left\{ 
{a^{\mbox{\dag }}(\mathbf{k})a(\mathbf{k})+\textstyle{1 \over 
2}+b^{\mbox{\dag }}(\mathbf{k})b(\mathbf{k})+\textstyle{1 \over 2}\;} 
\right\}
\end{equation}
\subsection{Dirac and Proca Supplemental Solutions}
\label{subsec:dirac}
As should be expected, the above analysis has its analogues for spin 1/2 and 
spin 1 fields. The Proca equation is so closely related to the Klein-Gordon 
equation that all results of the preceding sections can be readily 
extrapolated to spin 1 fields.

Since we do not have a classical spinor-field theory, one cannot obtain 
quantum spinor-field theory by quantizing it. The typical approach for 
spinors is to assume coefficient anti-commutation relations, in contrast to 
the coefficient commutation relations used for bosons. Doing so for 
supplemental fields, we obtain
\begin{equation}
\label{eq14}
\left[ {\underline{c}_{r} \left( {{\rm {\bf p}}} 
\right),\underline{c}_{s}^{\mbox{\dag }} \left( {{\rm {\bf {p}'}}} \right)} 
\right]_{+} =\left[ {\underline{d}_{r} \left( {{\rm {\bf p}}} 
\right),\underline{d}_{s}^{\mbox{\dag }} \left( {{\rm {\bf {p}'}}} \right)} 
\right]_{+} =-\delta_{rs} \delta_{{\rm {\bf p{p}'}}} \quad .
\end{equation}
Using (\ref{eq14}) in the free Hamiltonian density for Dirac supplemental particles 
results in a Hamiltonian analogous to (\ref{eq12}), and having 1/2 terms of opposite 
sign from those in the traditional theory.

\subsection{Supplemental Operators, Propagators, and Observables}
\label{subsec:supplemental}
Supplemental field number operators, Hamiltonian, creation/destruction 
operators, other observables, and propagators are derived in ref. 
\cite{Klauber:1} following steps parallel to those in the development 
of traditional QFT. These are (where for simplicity we ignore the 1/2 quanta 
terms in \underline {\textit{H}})
\begin{equation}
\label{eq15}
\begin{array}{l}
 \underline{a}^{\mbox{\dag }}\left( {{\rm {\bf k}}} 
\right)\underline{a}\left( {{\rm {\bf k}}} \right)\vert \underline{n}_{{\rm 
{\bf k}}} \rangle =\underline{N}_{a} \left( {{\rm {\bf k}}} \right)\vert 
\underline{n}_{{\rm {\bf k}}} \rangle =\underline{n}_{{\rm {\bf k}}} \vert 
\underline{n}_{{\rm {\bf k}}} \rangle =-\left| {\underline{n}_{{\rm {\bf 
k}}} } \right|\vert \underline{n}_{{\rm {\bf k}}} \rangle 
\,\,\,\,\,\,\,\,\,\,\,\,\,\,\,\,\underline{n}_{{\rm {\bf k}}} \le 0\, \\ 
 \underline{b}^{\mbox{\dag }}\left( {{\rm {\bf k}}} 
\right)\underline{b}\left( {{\rm {\bf k}}} \right)\vert 
\underline{\bar{{n}}}_{{\rm {\bf k}}} \rangle \,=\underline{N}_{b} \left( 
{{\rm {\bf k}}} \right)\vert \underline{\bar{{n}}}_{{\rm {\bf k}}} \rangle 
=\underline{\bar{{n}}}_{{\rm {\bf k}}} \vert \underline{\bar{{n}}}_{{\rm 
{\bf k}}} \rangle =-\left| {\underline{\bar{{n}}}_{{\rm {\bf k}}} } 
\right|\vert \underline{\bar{{n}}}_{{\rm {\bf k}}} \rangle 
\,\,\,\,\,\,\,\,\,\,\,\,\,\,\underline{\bar{{n}}}_{{\rm {\bf k}}} \le 0\,\, 
\\ 
 \end{array}
\end{equation}
\begin{equation}
\label{eq16}
\begin{array}{c}
 \underline{H}\vert \underline{n}_{{\rm {\bf k}}} \rangle =\sum\limits_{{\rm 
{\bf {k}'}}} {\omega_{{\rm {\bf {k}'}}} \left\{ {\underline{a}^{\mbox{\dag 
}}\left( {{\rm {\bf {k}'}}} \right)\underline{a}\left( {{\rm {\bf {k}'}}} 
\right)\;\,+\,\;\underline{b}^{\mbox{\dag }}\left( {{\rm {\bf {k}'}}} 
\right)\underline{b}\left( {{\rm {\bf {k}'}}} \right)} \right\}} \,\vert 
\underline{n}_{{\rm {\bf k}}} \rangle_{_{_{_{_{_{_{ } } } } } } } \\ 
 =\sum\limits_{{\rm {\bf {k}'}}} {\omega_{{\rm {\bf {k}'}}} \left\{ 
{\underline{N}_{a} \left( {{\rm {\bf {k}'}}} 
\right)+\underline{N}_{b} \left( {{\rm {\bf {k}'}}} \right)} 
\right\}\,} \vert \underline{n}_{{\rm {\bf k}}} \rangle =\underline{n}_{{\rm 
{\bf k}}} \omega_{{\rm {\bf k}}} \vert \underline{n}_{{\rm {\bf k}}} 
\rangle =-\left| {\underline{n}_{{\rm {\bf k}}} } \right|\omega_{{\rm {\bf 
k}}} \vert \underline{n}_{{\rm {\bf k}}} \rangle \;\; \\ 
 \end{array}
\end{equation}
\begin{equation}
\label{eq17}
\begin{array}{l}
 {\begin{array}{*{20}c}
 {\mbox{lowering\, operator\, increases}} \hfill \\
 {\mbox{\, }n_{{\rm {\bf k}}} \mbox{\, by\, }1\,\,\left( {\mbox{one\, less\, 
particle}} \right)} \hfill \\
\end{array} }\,\,\,\,\,\,\,\,\,\,\,\,\,\underline{a}\left( {{\rm {\bf k}}} 
\right)\vert \underline{n}_{{\rm {\bf k}}} \rangle =\sqrt 
{\underline{n}_{{\rm {\bf k}}} } \vert \underline{n}_{{\rm {\bf k}}} 
+1\rangle_{{\begin{array}{l}
 \\ 
 \\ 
 \end{array}}} \\ 
 {\begin{array}{*{20}c}
 {\mbox{raising\, operator\, decreases}} \hfill \\
 {\mbox{\, }n_{{\rm {\bf k}}} \mbox{\, by\, }1\,\left( {\mbox{one\, more\, 
particle}} \right)} \hfill \\
\end{array} }\,\,\,\,\,\,\,\,\,\,\,\,\,\,\underline{a}^{\mbox{\dag }}\left( 
{{\rm {\bf k}}} \right)\vert \underline{n}_{{\rm {\bf k}}} +1\rangle =\sqrt 
{\underline{n}_{{\rm {\bf k}}} } \vert \underline{n}_{{\rm {\bf k}}} \rangle 
\\ 
 \end{array}
\end{equation}
\begin{equation}
\label{eq18}
{\begin{array}{*{20}c}
 {\mbox{3-momentum}} \hfill \\
 {\mbox{of\, a\, single\, supplemental\, particle}} \hfill \\
\end{array} }\,\,\mbox{\, \, \, \, \, }\,\,\,\,\,\,\,\,\mbox{\, \, \, }{\rm 
{\bf P}}^{oper} \vert \underline{n}_{{\rm {\bf k}}} =-1\rangle ={\rm {\bf 
k}}\vert \underline{n}_{{\rm {\bf k}}} =-1\rangle 
\end{equation}
\begin{equation}
\label{eq19}
{\begin{array}{*{20}c}
 {\mbox{probability\, current\, density\, (}\propto \mbox{\, velocity)}} 
\hfill \\
 {\mbox{of\, a\, single\, supplemental\, particle}} \hfill \\
\end{array} }\,\,\,\,\,\,\,{\rm {\bf j}}^{oper} \vert \underline{n}_{{\rm 
{\bf k}}} =-1\rangle =-\,\frac{{\rm {\bf k}}}{V\omega_{{\rm {\bf k}}} 
}\,\vert \underline{n}_{{\rm {\bf k}}} =-1\rangle \,\,\,\,\,\,\,
\end{equation}
\begin{equation}
\label{eq20}
\mbox{pressure}\,\,\,\,\,\,\,\,\,\,\,\,\,\,\,\,\,\,\,\,\,\,\,\,\,\,\,\,\,\,\,\,\,\,\,\,\,\,\,\,\,\,\,\,\,\,\,\mbox{\, 
\, }T_{11} \vert \underline{n}_{{\rm {\bf k}}_{1} } \rangle =\textstyle{1 
\over V}\underline{n}_{{\rm {\bf k}}_{1} } \frac{\left( {k_{1} } 
\right)^{2}}{\omega_{{\rm {\bf k}}_{1} } }\vert \underline{n}_{{\rm {\bf 
k}}_{1} } \rangle =-\textstyle{1 \over V}\left| {\underline{n}_{{\rm {\bf 
k}}_{1} } } \right|\frac{\left( {k_{1} } \right)^{2}}{\omega_{{\rm {\bf 
k}}_{1} } }\vert \underline{n}_{{\rm {\bf k}}_{1} } \rangle 
\end{equation}
\smallskip
\begin{equation}
\label{eq21}
\mbox{Feynman\, propagator\, \, \, \, \, 
}\,\,\,\,\,\,\,\,\,\,\,\,\,\,\,\,\,\,\,\,\,\,\,\,\,\,\mbox{\, \, \, }{\Delta 
}_{F} \left( k \right)=-\Delta_{F} \left( k \right)
\end{equation}
Number operators (\ref{eq15}) yield number eigenvalues of opposite sign (negative) 
from their traditional counterparts. The energy (\ref{eq16}) of a supplemental 
particle state is negative. Supplemental propagators (\ref{eq21}) have the same 
form, but opposite sign from traditional propagators. And the total 
three-momentum (\ref{eq18}) for a supplemental particle state is in the opposite 
direction of its velocity (\ref{eq19}). Such characteristics, particularly the 
lattermost, are not those of real particles in our universe, though they can 
be (and often are) so for the virtual particles of traditional QFT. 

For example, the virtual exchange between two oppositely charged macroscopic 
bodies traveling the same line of action must entail three-momentum in the 
opposite direction of travel of the virtual particles in order for the 
bodies to attract. And virtual loop diagrams are integrated over both 
positive and negative energies for the individual virtual particles therein. 
Further, traditional scalar (timelike polarization) virtual photons have 
negative energies.\cite{Mandl:1984} Still further, traditional fermion 
zero-point energies are negative.

\subsection{Nature of Supplemental Particles}
\label{subsec:nature}
Hence, we consider herein that if supplemental particles are indeed realized 
in the spectrum of states, they are necessarily constrained to be virtual, 
cannot be real, and are never directly observed. No symmetry breaking 
mechanism (at least none much above contemporary energy levels) is 
envisioned (in the present version of the model) between the traditional and supplemental fields.

And to be clear, the supplemental particles, though having negative-energy 
states, are not a reincarnation of Dirac's sea of negative energy, but quite 
a different thing entirely. Neither are they related to the Wheeler-Feynman 
absorber theory, which is compared to the present theory in ref. 
\cite{Klauber:1}.

Note that if supplemental particles were to interact with traditional 
particles, then for every traditional Feynman diagram with a virtual 
particle represented by a traditional propagator, we would have another 
diagram, the same in every regard, except that a supplemental virtual 
particle would replace the traditional one. From (\ref{eq21}), this would result in 
only a sign change in the Feynman sub-amplitude for the interaction. When we 
add the two diagrams, which we need to do if the incoming and outgoing 
particles are the same, they cancel, leaving zero for the transition 
amplitude.

Thus, we would have zero probability for any interaction to take place, and 
nothing would happen in the world. Hence, we posit that for standard model 
gauge interactions, the traditional and supplemental particles are 
uncoupled. Quantum gravitational interactions are considerably more complex, 
however, and interactions of some sort between traditional and supplemental 
particles may occur.

\section{Potential Resolutions of Certain Extant Problems}
\label{potential}
\subsection{Cancellation of Zero-Point Energy Fluctuations}
\label{subsec:cancellation}
Weinberg\cite{Weinberg:1989} and Klauber\cite{Klauber:2013} (pg. 279), 
among others, show summing (or integration) of the zero-point energies of a 
boson field of mass $m$ up to a wave number cutoff $k_{c} \gg m$ on the order 
of the Planck scale yields a vacuum energy density exceeding the observed 
value by a factor of more than 120 orders of magnitude. 
Martin\cite{Martin:2012}, citing Zeldovich's\cite{Zel:1968} 
original work, notes that such an approach, though widely disseminated, is 
not valid, as it violates Lorentz invariance. That is, the cutoff wave 
number momentum is observer dependent, and just as the cutoff method fails 
for that reason in renormalization, it fails here, as well.

Using dimensional regularization (which is Lorentz covariant), instead of 
the cutoff method, to evaluate the integration over zero-point wave numbers, 
Martin finds a vacuum energy density differing from the observed value by an 
order of ``only'' 55 orders of magnitude. He also shows, via the same 
approach, a ZPE pressure of equal magnitude (in natural units) of, but 
opposite sign from, the ZPE energy density, i.e., an equation of state $w  =$ 
-1, which, of course, is that of a cosmological constant.

The Zeldovich/Martin approach nevertheless leaves us with the same 
qualitative problem - a mind bending discrepancy in magnitude between theory 
and observation.

However, if we consider the total (free, scalar) Hamiltonian as the sum of 
(\ref{eq12}) and (\ref{eq13}),
\begin{equation}
\label{eq22}
H_{\mbox{tot}} =H+\underline{H}=\sum\limits_{{\rm {\bf k}}} {\omega_{{\rm 
{\bf k}}} } \left\{ {a^{\dag }(\mathbf{k})a(\mathbf{k})+b^{\dag 
}(\mathbf{k})b(\mathbf{k})+\underline{a}^{\dag 
}(\mathbf{k})\underline{a}(\mathbf{k})+\underline{b}^{\dag 
}(\mathbf{k})\underline{b}(\mathbf{k})} \right\},
\end{equation}
then the 1/2 terms all drop out and the expectation energy of the vacuum 
from the free Hamiltonian is naturally zero. In similar fashion, the 
concomitant expectation value of pressure is likewise zero. In this 
scenario, the vacuum stress energy tensor is zero, i.e., the associated 
cosmological constant is zero.

\subsection{Cancellation of the Higgs Condensate Energy}
\label{subsec:condensate}
In electroweak symmetry breaking\cite{Klauber:2017}, the potential, 
expressed in terms of the Higgs field doublet $\Phi $, is
\begin{equation}
\label{eq23}
{\cal V}=\mu^{2}\Phi^{\mbox{\dag }}\Phi +\lambda \left( {\Phi^{\mbox{\dag 
}}\Phi } \right)^{2}.
\end{equation}
For the unitary gauge, the minimum value of the potential (\ref{eq23}), where the 
Higgs field takes the VEV $v$, we find a Higgs condensate energy density with 
VEV
\begin{equation}
\label{eq24}
\overline {\cal V}_{\min } =-\frac{\lambda 
v^{4}}{4}\,\,\,\,\,\,\,\,\,\,\,\,\mbox{where}\,\,\,\,\,v=\sqrt {\frac{-\mu 
^{2}}{\lambda }} \,\,\,\,\,\,\,\,\,\mu^{2}\,<\,\,0,\,\,\,\lambda >0\quad .
\end{equation}
As noted in ref. \cite{Klauber:2017}, the relation (\ref{eq23}) does not really 
make sense numerically, because $\Phi $ is an operator that creates and 
destroys fields. $\Phi $ does not take on different values, so a numerical 
quantity, such as ${\cal V}$ is often presumed to be in (\ref{eq23}), cannot be 
functionally dependent on it. What (\ref{eq23}) implies is an expectation value for 
the potential
\begin{equation}
\label{eq25}
\begin{array}{c}
 \overline {\cal V} =\langle n_{{\rm {\bf k}}} ,n_{{\rm {\bf {k}'}}} 
...\vert {\cal V}\vert n_{{\rm {\bf k}}} ,n_{{\rm {\bf {k}'}}} ...\rangle \\ 
 =\mu^{2}\langle n_{{\rm {\bf k}}} ,n_{{\rm {\bf {k}'}}} ...\vert \Phi 
^{\mbox{\dag }}\Phi \vert n_{{\rm {\bf k}}} ,n_{{\rm {\bf {k}'}}} ...\rangle 
+\lambda \langle n_{{\rm {\bf k}}} ,n_{{\rm {\bf {k}'}}} ...\vert \left( 
{\Phi^{\mbox{\dag }}\Phi } \right)^{2}\vert n_{{\rm {\bf k}}} ,n_{{\rm {\bf 
{k}'}}} ...\rangle\, \\ 
 \end{array}
\end{equation}
Given the operator nature of $\Phi $, (\ref{eq25}) reduces to a dependence on the 
number densities
\begin{equation}
\label{eq26}
\overline {\cal V} =\mu^{2}\frac{1}{V}\left( {n_{{\rm {\bf k}}} +n_{{\rm 
{\bf {k}'}}} +...} \right)+\lambda \frac{1}{V^{2}}\left( {n_{{\rm {\bf 
k}}}^{2} +n_{{\rm {\bf {k}'}}}^{2} +...} \right).
\end{equation}
Thus, the famous Mexican hat diagram corresponds to the expectation value of 
the potential on the vertical axis, and the number density of Higgs particle 
states on the horizontal axes.

Hence, the relations (\ref{eq24}) are actually deduced in a framework where number 
densities are positive, i.e., $n_{\mathbf{k}}$ \textgreater 0. However, a 
parallel analysis, with the same algebraic form (\ref{eq23}) for supplemental 
particles, would yield
\begin{equation}
\label{eq27}
\begin{array}{c}
 \,\underline{\overline {\cal V} }=\mu^{2}\frac{1}{V}\left( 
{\underline{n}_{{\rm {\bf k}}} +\underline{n}_{{\rm {\bf {k}'}}} +...} 
\right)+\lambda \frac{1}{V^{2}}\left( {\underline{n}_{{\rm {\bf k}}}^{2} 
+\underline{n}_{{\rm {\bf {k}'}}}^{2} +...} 
\right)_{_{_{_{_{_{{\begin{array}{l}
 \\ 
 
 \end{array}}} } } } } } \\ 
 \,\,\,\,\,\,\,\,\,\,\,\,\,\,\,\,\,\,=-\mu^{2}\frac{1}{V}\left( {\left| 
{\underline{n}_{{\rm {\bf k}}} } \right|+\left| {\underline{n}_{{\rm {\bf 
{k}'}}} } \right|+...} \right)+\lambda \frac{1}{V^{2}}\left( 
{\underline{n}_{{\rm {\bf k}}}^{2} +\underline{n}_{{\rm {\bf {k}'}}}^{2} 
+...} \right)\,. \\ 
 \end{array}
\end{equation}
Finding the minimum of (\ref{eq27}), where the supplemental field \underline {}$\Phi 
$ takes the VEV value $v$, implies, symbolically and parallel to (\ref{eq23}), that 
effectively, the sign has changed in the $\mu^{2}$ term, i.e.,
\begin{equation}
\label{eq28}
\underline{{\cal V}}=-\mu^{2}\underline{\Phi }^{\mbox{\dag 
}}\underline{\Phi }+\lambda \left( {\underline{\Phi }^{\mbox{\dag 
}}\underline{\Phi }} \right)^{2}.
\end{equation}
And again, in parallel fashion, finding the stationary (maximum, in this 
case [see Section \ref{subsec:mylabel2}]) value of (\ref{eq28}) yields
\begin{equation}
\label{eq29}
\,\underline{\overline {\cal V} }_{\max } =-\frac{\lambda 
v^{4}}{4}\,=\frac{\left| \lambda 
\right|v^{4}}{4}\,\,\,\,\,\,\,\,\,\,\,\,\,\,\,\mbox{where}\,\,\,\,\,v=\sqrt 
{\frac{\mu^{2}}{\lambda }} \,\,\,\,\,\,\,\,\,\,\,\mu 
^{2}\,<\,\,0,\,\,\,\lambda <0.
\end{equation}
Adding the traditional Higgs energy VEV (\ref{eq24}) to the supplemental Higgs value 
(\ref{eq29}) yields zero, which matches observational constraints.

\subsection{Gauge Hierarchy and Supplemental Solutions}
\label{subsec:gauge}
As noted in (\ref{eq21}), supplemental field propagators have opposite sign from, 
but equal magnitude of, their sibling traditional field propagators. Thus, 
if a single propagator in a loop of a problematic transition sub-amplitude 
with traditional fields were replaced by its sibling supplemental 
propagator, the resulting transition sub-amplitude would have opposite sign 
from, but equal magnitude of, the original sub-amplitude. And adding them 
would yield zero contribution to the total amplitude.

To this end, consider the traditional Higgs-lepton Lagrangian at the false 
vacuum,
\begin{equation}
\label{eq30}
\,\,{\cal L}^{\,\,LH}=-g_{l} \left( {\bar{{\Psi }}_{l}^{L} \psi_{l}^{R} 
\Phi +\Phi^{\mbox{\dag }}\bar{{\psi }}_{l}^{R} \Psi_{l}^{L} } 
\right)-g_{\nu_{l} } \left( {\bar{{\Psi }}_{\nu_{l} }^{L} \psi_{\nu_{l} 
}^{R} \tilde{{\Phi }}+\tilde{{\Phi }}^{\mbox{\dag }}\bar{{\psi }}_{\nu_{l} 
}^{R} \Psi_{\nu_{l} }^{L} } \right)
\end{equation}
and make the replacement
\begin{equation}
\label{eq31}
\Psi_{l}^{L} \to \Psi_{l}^{L} +\underline{\Psi }_{l}^{L} 
\,\,\,\,\,\,\,\,\,\,\,\,\,\,\,\,\,\,\,\,\,\psi_{l}^{R} \to \psi_{l}^{R} 
+\underline{\psi }_{l}^{R} ,
\end{equation}
to get (where for simplicity we ignore the neutrino terms in $v_{l})$
\begin{equation}
\label{eq32}
\begin{array}{c}
 \,\,{\cal L}_{l}^{\,\,LH} =-g_{l} \left( {\left( {\overline \Psi_{l}^{L} 
+\underline{\overline \Psi }_{l}^{L} } \right)\left( {\psi_{l}^{R} 
+\underline{\psi }_{l}^{R} } \right)\Phi +\Phi^{\mbox{\dag }}\left( 
{\bar{{\psi }}_{l}^{R} +\underline{\overline \psi }_{l}^{R} } \right)\left( 
{\Psi_{l}^{L} +\underline{\Psi }_{l}^{L} } \right)} \right) \\ 
 =-g_{l} \bar{{\Psi }}_{l}^{L} \psi_{l}^{R} \Phi -g_{l} \bar{{\Psi 
}}_{l}^{L} \underline{\psi }_{l}^{R} \Phi -g_{l} \underline{\overline \Psi 
}_{l}^{L} \psi_{l}^{R} \Phi -g_{l} \underline{\overline \Psi }_{l}^{L} 
\underline{\psi }_{l}^{R} \Phi \,\,\,+\,\,\,h.c. \\ 
 \end{array}
\end{equation}

\begin{figure}[htbp]
\centerline{\includegraphics[width=4.80in,height=0.72in]{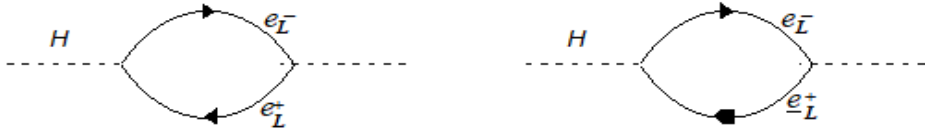}}
\caption{Traditional (LHS) vs Supplemental (RHS) Higgs Mass Radiative 
Corrections}
\label{fig1}
\end{figure}

Focusing on terms in (\ref{eq32}) where $l=$1 (i.e., only the electron family), 
consider the Higgs mass radiative correction interaction of 
Figure 1 where in the RH sub-diagram, we introduce 
a pentagon symbol to represent a supplemental particle trajectory and 
underlining for supplemental particles themselves.

The LH vertex in the LHS sub-diagram of Figure 1 arises from a traditional 
radiative correction term, the first term in second row of (\ref{eq32}). The RH 
vertex in the same sub-diagram arises from the h.c. term in (\ref{eq32}) 
corresponding to the first term.

Now consider the RHS sub-diagram, which has a supplemental positron 
propagator in place of the traditional positron propagator of the LHS 
sub-diagram. The LH vertex of the RHS sub-diagram arises from the second 
term in the second row of (\ref{eq32}); the RH vertex, from the associated h.c. 
term.

The two sub-diagrams in Figure 1 have equal magnitude, but due to the sign 
reversal for supplemental propagators (\ref{eq21}), they have Feynman amplitudes of 
opposite sign. They cancel.

Extending the analysis to other leptons and quarks, and to other relevant 
sub-diagrams, one finds the same cancellation. From this perspective, there 
are no radiative corrections to the Higgs mass.

Note that from the similarity between the first and fourth terms in the 
bottom row of (\ref{eq32}), when the Higgs symmetry breaks, supplemental fields 
obtain masses with identical magnitudes of their traditional brethren.

Note, further, it is presumed that traditional and supplemental particles do not 
couple to the vector bosons of the standard model (SM). If they did, then 
the same type of cancellation would occur, and we would not see the running 
SM coupling constants observed in experiment. In fact, as noted earlier, 
transition sub-amplitudes would all cancel with one another, probability for 
anything to happen would be zero, and we would have no phenomenological 
universe of interacting particles at all.

\section{Resolution of Seeming Inconsistencies}
\label{sec:resolution}
\subsection{Non-Positive Definite Metric}
\label{subsec:non-positive}
As shown in ref. \cite{Klauber:1}, supplemental eigenstate norms are 
not positive 
definite\cite{Kall:1972}\cite{Jauch:1976}\cite{Gupta:1977}. 
Specifically, for suitable normalization, where numbers refer to \underline 
{\textit{a}}(\textbf{k})-type particles [having negative particle numbers] 
of the same \textbf{k},
\begin{equation}
\label{eq33}
\left\langle {0} \mathrel{\left| {\vphantom {0 0}} \right. 
\kern-\nulldelimiterspace} {0} \right\rangle =1\,\,\,\,\,\,\,\,\left\langle 
{-1} \mathrel{\left| {\vphantom {{-1} {-1}}} \right. 
\kern-\nulldelimiterspace} {-1} \right\rangle =-1\,\,\,\,\,\,\,\left\langle 
{-2} \mathrel{\left| {\vphantom {{-2} {-2}}} \right. 
\kern-\nulldelimiterspace} {-2} \right\rangle 
=1\,\,\,\,\,\,\,\,\,\left\langle {-3} \mathrel{\left| {\vphantom {{-3} 
{-3}}} \right. \kern-\nulldelimiterspace} {-3} \right\rangle 
=-1\,\,\,\,\,\,\,............
\end{equation}
So, the non-positive definite metric in Fock space for such particles is
\begin{equation}
\label{eq34}
\underline{g}_{mn}^{Fock} =\left[ {{\begin{array}{*{20}c}
 1 \hfill & 0 \hfill & 0 \hfill & 0 \hfill & {..} \hfill \\
 0 \hfill & {-1} \hfill & 0 \hfill & 0 \hfill & {..} \hfill \\
 0 \hfill & 0 \hfill & 1 \hfill & 0 \hfill & {..} \hfill \\
 0 \hfill & 0 \hfill & 0 \hfill & {-1} \hfill & {..} \hfill \\
 {..} \hfill & {..} \hfill & {..} \hfill & {..} \hfill & {...} \hfill \\
\end{array} }} \right],
\end{equation}
This result differs from the identity matrix of the traditional-particle 
Fock space metric. Pauli\cite{Pauli:1941}\cite{Dirac:1942}\cite{Pauli:1943}
investigated fields with commutation relations such as (\ref{eq10}), which generate 
metrics such as (\ref{eq34}), and concluded it was impossible for them to be 
realized in the spectrum of physical states. In pre-QFT quantum theory, this 
would imply negative probabilities for some states, and positive ones for 
others.

However, if supplemental particles are only virtual and never real, then 
their metric in Fock space should be irrelevant, as that metric is a 
relationship between real states. Nevertheless, we can resolve this issue, 
even if such particles were real, as follows.

In QFT, one focuses on expectation values of operators/observables. In the 
traditional theory, for an eigenstate $\vert n_{{\rm {\bf k}}} \rangle $ 
having an eigenvalue ${\mathit{o}}_{{\rm {\bf k}}} $ of an operator ${\cal O}$,
\begin{equation}
\label{eq35}
\overline {\cal O} \equiv \langle n_{{\rm {\bf k}}} \vert 
{\cal O}\,\,\vert n_{{\rm {\bf k}}} \rangle ={\mathit{o}}_{{\rm {\bf k}}} 
\left\langle {n_{{\rm {\bf k}}} } \mathrel{\left| {\vphantom {{n_{{\rm {\bf 
k}}} } {n_{{\rm {\bf k}}} }}} \right. \kern-\nulldelimiterspace} {n_{{\rm 
{\bf k}}} } \right\rangle ={\mathit{o}}_{{\rm {\bf k}}} .
\end{equation}
Directly extrapolating (\ref{eq35}) to supplemental particles, one gets 
\begin{equation}
\label{eq36}
\overline {\cal O} =\langle \underline{n}_{{\rm {\bf k}}} 
\vert {\cal O}\vert \underline{n}_{{\rm {\bf k}}} \rangle =\underline{{\mathit{o}}}_{{\rm {\bf k}}} \left\langle {\underline{n}_{{\rm {\bf k}}} } 
\mathrel{\left| {\vphantom {{\underline{n}_{{\rm {\bf k}}} } 
{\underline{n}_{{\rm {\bf k}}} }}} \right. \kern-\nulldelimiterspace} 
{\underline{n}_{{\rm {\bf k}}} } \right\rangle =\underline{{\mathit{o}}}_{{\rm 
{\bf k}}} \left( {-1} \right)^{\underline{n}_{{\rm {\bf k}}} },
\end{equation}
a not very satisfying, consistent, or usable result, as our expectation 
value is positive for every other multiparticle eigenstate and negative for 
the ones in between.

However, (\ref{eq35}) was defined in a manner that would give us a numerical value 
corresponding to what we would measure for traditional particles. It is a 
simple and straightforward matter to define a mathematical procedure 
slightly different from (\ref{eq36}) that will give us what we would measure for 
supplemental particle states, i.e., in the present case, the eigenvalue 
(with correct sign). Thus, we define expectation values for supplemental 
eigenstates as
\begin{equation}
\label{eq37}
\overline {\cal O} \equiv \left( {-1} 
\right)^{\underline{n}_{{\rm {\bf k}}} }\langle \underline{n}_{{\rm {\bf 
k}}} \vert {\cal O}\vert \underline{n}_{{\rm {\bf k}}} \rangle =\left( {-1} 
\right)^{\underline{n}_{{\rm {\bf k}}} }\underline{{\mathit{o}}}_{{\rm {\bf k}}} 
\left\langle {\underline{n}_{{\rm {\bf k}}} } \mathrel{\left| {\vphantom 
{{\underline{n}_{{\rm {\bf k}}} } {\underline{n}_{{\rm {\bf k}}} }}} \right. 
\kern-\nulldelimiterspace} {\underline{n}_{{\rm {\bf k}}} } \right\rangle 
=\underline{{\mathit{o}}}_{{\rm {\bf k}}} \left( {-1} 
\right)^{2\underline{n}_{{\rm {\bf k}}} }=\underline{{\mathit{o}}}_{{\rm {\bf 
k}}} .
\end{equation}
Hence, any physical quantity we might wish to measure related to 
supplemental particles can be determined theoretically via (\ref{eq37}), and the 
impasse that stymied Pauli dissolves. This definition would also apply to 
multi-particle states (with a factor of $\left( {-1} 
\right)^{\underline{n}_{{\rm {\bf k}}} }$ for each \textbf{k} and similar 
factors for supplemental \underline {\textit{b}}$_{\mathbf{k}}$-type 
antiparticles) and transition amplitudes. 

(\ref{eq35}) and (\ref{eq37}) can be generalized using the coefficient commutation relations 
(\ref{eq10}) along with their traditional counterparts for traditional fields, where 
a bar through the middle of a quantity indicates it can be for either a 
traditional or supplemental field.
\begin{equation}
\label{eq38}
\overline {\cal O} \equiv \left[ {\rlap{--} {a}_{{\rm {\bf 
k}}} ,\rlap{--} {a}_{{\rm {\bf k}}}^{\mbox{\dag }} } \right]^{\rlap{--} 
{n}_{{\rm {\bf k}}} }\langle \rlap{--} {n}_{{\rm {\bf k}}} \vert {\cal 
O}\vert \rlap{--} {n}_{{\rm {\bf k}}} \rangle =\rlap{--} {{\mathit{o}}}_{{\rm 
{\bf k}}} 
\end{equation}
\subsection{Vacuum Decay}
\label{subsec:vacuum}
Linde\cite{Linde:1984}\cite{Linde:1988} conjectured ``some 
mechanism, associated probably with some kind of symmetry of the elementary 
particle theory, which would automatically lead to a vanishing of the 
cosmological constant in a wide class of theories''. He considered the 
possibility of two quasi-independent universes, occupying the same physical 
space and having the same particle equations of motion, but with opposite 
signs for their Lagrangians, for which he coined the term ``antipodal 
symmetry''. The two universes would not interact via standard model forces, 
but would be linked gravitationally. Due to its negative Lagrangian, the 
vacuum energy of the antipodal (shadow) universe would be of opposite sign 
from ours, so the total vacuum energy would net to zero. Linde's approach 
and the one shown herein have some similarities, but are quite different at 
their roots. 

However, in personal correspondence with the present author, Linde noted his 
dissatisfaction with his idea as it seemed inevitably to imply vacuum 
instability and decay. Interaction between the two types of particles, even 
if only gravitationally, would lead to instability of the vacuum. In 
essence, a positive-energy traditional particle and a negative-energy 
supplemental particle could arise spontaneously from the vacuum, as they 
would comprise a net total energy of zero.

\begin{figure}[htbp]
\centerline{\includegraphics[width=3.90in,height=0.80in]{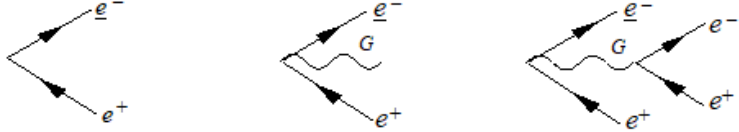}}
\caption{Conjectured Modes of Vacuum Decay Prohibited in One Form of Theory}
\label{fig2}
\end{figure}

As supplemental particles and traditional particles are presumed to be 
prohibited from interacting via standard model forces, the interactions 
displayed in Figure 2 are solely gravitational. The 
$G$ symbol represents a traditional graviton, but in possible embodiments of 
supplemental field theory, it could also represent a supplemental graviton.

Further, with reference to Section 4.3, the $G$ in the 
figure could be replaced by an $H$, i.e., the Higgs field could also interact 
with leptons and quarks in the manner shown in 
Figure 2.

At energies above the weak symmetry breaking level, the first diagram in 
Figure 2 is prohibited, because no terms of form $\underline{\bar{{\psi 
}}}^{\mbox{\dag }}\psi $ are anticipated in the Lagrangian. However, when 
the Higgs gets a VEV, the second and third terms in the bottom row of (\ref{eq32}) 
yield Lagrangian terms represented by that diagram and similar ones.

However, for that and the other diagrams in the figure, via
the form of the theory suggested herein whereby real (on-shell) 
supplemental particles are prohibited, no interaction leaving a final state 
with real supplemental particles can occur. So, none of the interactions of 
Figure 2 would be allowable, even though they 
maintain energy and momentum balance. Thus, the vacuum could not decay in a 
manner like those of the diagrams in Figure 2, and 
it appears that for this interpretation of supplemental particles, Linde's 
concern is resolved.

\subsection{Supplemental Energy Bounded from Above}
\label{subsec:mylabel2}
Resistance to negative-energy particles can also arise with regard to the 
fundamental physical principle that configurations of particles 
spontaneously seek the lowest attainable energy state. Thus, negative-energy 
particles would have no lowest state and be inherently unstable.

\subsubsection{For Supplemental Particles Constrained to be Virtual Only}
\label{subsubsec:For}
If supplemental particles can never be real, but only virtual, this concern 
is not relevant. There would be no real supplemental particles subject to 
such constraint. Virtual particles are not subject to it.

\subsubsection{If Supplemental Particles Could be Real}
\label{subsubsec:if}
If supplemental particles were real, it is wholly reasonable to presume they 
would seek the highest (least-negative) energy state. That is, with regard 
to energy levels, they would behave as traditional particles under a mirror 
reflection of the energy axis.

\section{Additional Implications}
\label{sec:Additional}
\subsection{Arrow of Time}
\label{subsec:Arrow}
Some, such as Barbour et al\cite{Barbour:2014}, and Carroll and 
Chen\cite{Carroll:2004}, have suggested the possible existence of a mirror 
universe to ours moving, since the Big Bang, in the opposite direction of 
time.

The absence of observed real supplemental particles may be conjectured as 
being related to the fact that reversing the arrow of time in the 
supplemental solutions produces the traditional solutions. That may mean, 
with regard to perception, that real supplemental particles travel backward 
in time. Thus, one might speculate that the universe needed no initial 
energy from which to begin, with equal amounts of traditional and 
supplemental particles emerging from nothing. From there, the real 
negative-energy supplemental particles traveled backward in time, and in the 
process created their own universe, which would appear to any beings in that 
universe to possess positive energy. The traditional particles, traveling 
forward in time, created our universe and appear to us as having positive 
energy. Each universe would have only one kind of real particle, but virtual 
particles of both types. And this would result in null (or near null, 
perhaps for other reasons) vacuum energy in both. 

\subsection{Inflation, GUTS, and More}
\label{subsec:Inflation}
Inflation, GUT, and possibly other vacuum condensate energies could cancel 
when account is taken of the associated supplemental fields in similar 
manner to that shown herein for the Higgs condensate.

\subsection{Dark Energy}
\label{subsec:dark}
A slight asymmetry, from some presently unknown cause, between traditional 
and supplemental fields could give rise, via ZPE or symmetry breaking 
mechanisms, to a small cosmological constant, similar to what is observed.

\section{Summary, Caveats, and Questions}
\subsection{Summary}
A previously unrecognized, but fundamental, symmetry in elementary particle 
theory exists in which supplemental (alternative) forms for the solutions to the QFT 
free-field equations are obtained from the traditional solutions by taking 
$\omega \to -\omega $. The incorporation of these alternative solutions into the 
theory results in a total Hamiltonian yielding null ZPE and possible 
resolutions of the Higgs hierarchy and condensate energy problems. Agreement 
with observations should be maintained if supplemental states occur only as 
virtual, and not as real, particles, and given the properties of the 
supplemental states, this appears reasonable.

The supplemental/traditional solution symmetry can be maintained over all 
energy scales, unlike other attempts to null out vacuum energy, such as 
supersymmetry, which only succeed down to particular, non-contemporary, 
energy levels. 

Resolution of certain significant issues in physics and cosmology is direct 
and simple. Nothing more, no new theory, is needed, other than the inclusion 
in QFT of alternative solution forms to the field equations.

\subsection{Caveats}
\label{subsec:caveats}
It may seem somewhat \textit{ad hoc} to simply allocate terms selectively to the Lagrangian 
that give a desired result, even though that is what physics has done from 
day one. In the present case, we have added Higgs-supplemental fields 
interaction and Higgs supplemental field potential terms. At present there 
is no hint of a more encompassing theory from which these terms may spring.

While Section \ref{subsec:condensate} presents a way in which the Higgs 
condensate energy density can be nulled out, it appears to contain an 
implicit assumption that supplemental Higgs particles are real. (The 
horizontal axes in the traditional and supplemental Mexican hat diagrams 
represent expectation values for real particle density.) But, a fundamental 
assumption in the version of the theory proposed herein, used to 
potentially resolve the Higgs hierarchy problem, and employed to avoid 
vacuum decay, was that supplemental particles could only be virtual, and not 
real. Perhaps, the Higgs condensate issue, and the stable true vacuum 
location determination, can still be applied in a sense for which the 
supplemental Higgs is purely virtual. Possibly the virtual interactions 
between the traditional Higgs and the supplemental Higgs result in an 
effective supplemental potential that tracks the traditional potential.

The author has not found any way, in the pure mathematics of the theory, to 
prohibit supplemental particles from manifesting as real particles. The 
presumption for constraining them to be virtual is based on the empirical 
reality that negative-energy particles with momentum in the opposite 
direction of velocity are simply not observed, and in fact, do not even make 
sense in our universe as we know it. However, this leads to a non-trivial 
issue regarding unitarity.

That is, in the traditional theory, the S matrix satisfies unitarity, i.e., 
\begin{equation}
\label{eq39}
\sum\limits_f {\left| {S_{fi} } \right|^{2}} =1\quad ,
\end{equation}
where \textbar $S_{fi}$\textbar $^{2}$ is the probability that a given 
initial state $i$ will transition into a particular final state $f$. 

After incorporating renormalization, (\ref{eq39}) arises naturally from the 
mathematics of the theory. For a given initial state $i$, all possible final 
states $f$ (all which are permitted mathematically) need to be included in 
(\ref{eq39}). For the initial state \textbar 0$\rangle $, this means, in the present 
version of the theory, final multiparticle states that include supplemental 
particles, such as those shown in Fig. 2, should be included in (\ref{eq39}). But we 
have prohibited them on physical (not mathematical) grounds.

The issue can be resolved, though perhaps in an \textit{ad hoc} manner, by normalizing the 
$S_{fi}$, much as we normalize state vectors in quantum mechanics after wave 
function collapse. That is, we take all $S_{fi}$ equal to zero that have one 
or more final supplemental particles. Then the new $S$ matrix (underbar) would 
satisfy
\begin{equation}
\label{eq40}
\sum\limits_{{f}'} {\left| {\underline{S}_{{f}'i} } \right|^{2}} 
=1\,\,\,\,\,\,\,\,\left| {\underline{S}_{fi} } 
\right|^{2}=\frac{\left| {S_{fi} } \right|^{2}}{\,\sum\limits_{{f}'} {\left| 
{S_{{f}'i} } \right|^{2}} }\,\,\,\,\,\,{f}'\,\mbox{\,=\, final\, 
states\, with\, no\, supplemental\, particles}.
\end{equation}
This, in some sense, may be considered unnatural, but perhaps not so much 
more so than renormalization in QFT or normalization after wave function 
collapse in quantum mechanics. Those procedures were incorporated into 
theory mathematics after the base theory itself failed to match physical 
reality.

There may be better ways to make supplemental theory unitary and avoid 
vacuum decay. The topic deserves further research by the present author and 
perhaps by interested others.

\subsection{Question}
\label{subsec:question}
The fundamental question is whether or not nature in her physical 
manifestation parallels the nature of her mathematics. Supplemental field 
mathematical expressions solve the field equations and give rise to
a unique class of particles, but do such supplemental 
particles actually exist? The intractable nature of several outstanding 
problems in physics may lend support to them in some form.

\section*{Acknowledgements}
I thank Stephen Kelley, Andrei Linde, and Robin Ticciati for reading 
versions of the arXiv manuscript\cite{Klauber:2007} that preceded this 
article, and for offering valuable insights, suggestions, and references. I 
further thank Luc Longtin, Christian Maennel, Holger Teutsch, and Robin 
Ticciati once again for 
discussions and comparably valuable feedback on the present manuscript.

\section*{Appendix : Supplemental Particles and Independence from the Traditional 
Particles}
Since the traditional independent solutions of the Klein-Gordon equation
\begin{equation}
\label{eq41}
\phi \,\,\,\,=\sum\limits_{{\rm {\bf k}}=-\infty }^{+\infty } 
{\frac{1}{\sqrt {2V\omega_{{\rm {\bf k}}} } }\,a\left( {{\rm {\bf k}}} 
\right)e^{-i\left( {\omega t-{\rm {\bf k}}\cdot {\rm {\bf x}}} \right)}} 
\,\,+\sum\limits_{{\rm {\bf k}}=-\infty }^{+\infty } {\frac{1}{\sqrt 
{2V\omega_{{\rm {\bf k}}} } }b^{\dag }\left( {{\rm {\bf k}}} 
\right)e^{+i\left( {\omega t-{\rm {\bf k}}\cdot {\rm {\bf x}}} \right)}} 
\,\,\,
\end{equation}
are summed over all \textbf{k}, one might reason, that for each $+\omega $ 
in the summation with a --$k_{x}$, there is a $+\omega $ with a 
$+k_{x}$, and similarly for --$\omega $, leading to the solution forms (\ref{eq4}) 
and (\ref{eq5}). And thus, since the supplemental solutions are not mathematically 
independent from the traditional ones, they are already included in the 
theory and have nothing new to add. We respond to this argument below.

\medskip
\noindent
Consideration 1

As demonstrated in this article, by simply assuming supplemental solutions 
having the form shown herein, and proceeding step-by-step through the 
development of QFT with these solutions included, one gets distinctly 
different results, which cannot be subsumed into the traditional theory. 
Specifically, states created and destroyed by supplemental field operators 
have distinctly different eigenvalues (quantum numbers) from traditional 
particle states, and states with different quantum numbers are independent 
states.

Supplemental particles may be thought of as simply a new species of 
particle, much like SUSY particles are a different species. SUSY and 
standard model field equation solutions are not linearly independent with 
regard to spacetime coordinates, but the associated operators are 
independent.

\medskip
\noindent
Consideration 2

Consider traditional RQM where the independent solutions to the Klein-Gordon 
equation are shown in the summations of (\ref{eq41}). Supplemental (or alternative) 
solutions, represented by underbars are, from (\ref{eq4}) and (\ref{eq5}), shown in the 
summations of
\begin{equation}
\label{eq42}
\underline{\phi }\underline{}=\sum\limits_{{\rm {\bf k}}} {\frac{1}{\sqrt 
{2V\omega_{{\rm {\bf k}}} } }\underline{a}\left( {{\rm {\bf k}}} 
\right)e^{i\left( {\omega t+{\rm {\bf k}}\cdot {\rm {\bf x}}} \right)}} 
\,\,+\,\,\,\sum\limits_{{\rm {\bf k}}} {\frac{1}{\sqrt {2V\omega_{{\rm {\bf 
k}}} } }\underline{b}^{\dag }\left( {{\rm {\bf k}}} \right)e^{-i\left( 
{\omega t+{\rm {\bf k}}\cdot {\rm {\bf x}}} \right)}} .
\end{equation}
Mathematically, the terms in (\ref{eq42}) are not independent of those in (\ref{eq41}), 
since
\begin{equation}
\label{eq43}
a\left( {{\rm {\bf k}}} \right)e^{-i\left( {\omega t-{\rm {\bf k}}\cdot {\rm 
{\bf x}}} \right)}\,=\,\,\underline{b}^{\dag }\left( {-{\rm {\bf k}}} 
\right)e^{-i\left( {\omega t+\left( {-{\rm {\bf k}}} \right)\cdot {\rm {\bf 
x}}} \right)}\,\,\,\,\,\,\,\,\,\,\,\,\,\,\,\mbox{for\, \, \, }a\left( {{\rm 
{\bf k}}} \right)=\underline{b}^{\dag }\left( {-{\rm {\bf k}}} \right).
\end{equation}
Physically, however, consider the phase velocity of a traditional solution
\begin{equation}
\label{eq44}
e^{-i\left( {\omega t-k_{1} x_{1} } 
\right)}\,\,\,\,\,\,\,\,\,\,\mathrel{\mathop{\kern0pt\longrightarrow}\limits_{\mbox{phase}}^{\mbox{constant}}} 
\,\,\,\,\,\,\,\omega t-k_{1} x_{1} =\mbox{constant}\,\,\,\,\,\to 
\,\,\,\,\,v_{phase} =\frac{dx_{1} }{dt}=\frac{\omega }{k_{1} }\,\,,
\end{equation}
and phase is in the direction of wave number $k_{1}$. For $k_{1}$ in the 
positive direction, the wave travels in the positive direction.

Then, consider the phase velocity of a supplemental (alternative) solution
\begin{equation}
\label{eq45}
e^{i\left( {\omega t+k_{1} x_{1} } 
\right)}\,\,\,\,\,\,\,\,\,\,\mathrel{\mathop{\kern0pt\longrightarrow}\limits_{\mbox{phase}}^{\mbox{constant}}} 
\,\,\,\,\,\,\,\omega t+k_{1} x_{1} =\mbox{constant}\,\,\,\,\,\to 
\,\,\,\,\,v_{phase} =\frac{dx_{1} }{dt}=-\frac{\omega }{k_{1} }\,\,,
\end{equation}
and phase is in the opposite direction of wave number $k_{1}$. For $k_{1}$ in 
the positive direction, the wave travels in the negative direction.

One could, of course, define wave number as being in the opposite direction 
of phase velocity and have a consistent theory. Wave number direction is 
arbitrarily defined.

However, if positive $k_{1}$ is proportional to positive momentum, as in
\begin{equation}
\label{eq46}
k_{1} =\frac{p_{1} }{\hbar},
\end{equation}
then traditional and supplemental solutions with momentum in the same 
direction have phase velocity in opposite directions. (This is also true for 
states in QFT, as shown in the main body of this article.)

If we developed our theory, using only (\ref{eq42}) and not (\ref{eq41}), and also revised 
our assumption in (\ref{eq46}) to
\begin{equation}
\label{eq47}
k_{1} =-\frac{p_{1} }{\hbar},
\end{equation}
we would find (\ref{eq45}) as
\begin{equation}
\label{eq48}
v_{phase} =\frac{dx_{1} }{dt}=-\frac{\omega }{k_{1} }=\frac{\hbar\omega 
}{p_{1} },
\end{equation}
and phase velocity and momentum would be in the same direction, as we find 
in the world.

One could then consider the difference between the two theories is in 
assumptions (\ref{eq46}) and (\ref{eq47}). As long as we use either (\ref{eq46}) with (\ref{eq41}) or (\ref{eq47}) 
with (\ref{eq42}), we get the same results. The choice between the two is 
essentially only a choice of gauge.

However, one could instead postulate a different species of particle 
governed by (\ref{eq47}) instead of (\ref{eq46}), but with solutions of form (\ref{eq41}). Or 
equivalently, one could get the same results by postulating that particle is 
governed by (\ref{eq46}) with solutions of form (\ref{eq42}).

In this article we do the latter, as it is simpler to develop, more suited 
to QFT, and seems more natural (since we do not change fundamental physical 
relationships between quantities, but merely re-cast the solution forms to 
the field equations.)

In essence, mathematically, the supplemental solutions are contained within 
the traditional solutions, but, physically, they give rise to particle 
states not found in the traditional theory, and in this context, may be 
considered independent.


\end{document}